\documentclass[10pt,preprint]{emulateapj-rtx4}
\usepackage{graphicx}
\usepackage{epstopdf}
\usepackage{pdflscape}

\newcommand{ \um}{$\mu$m~}
\newcommand{ \ums}{$\mu$m}
\def\kmsMpc{\ifmmode {\rm\,km\,s^{-1}\,Mpc^{-1}}\else
    ${\rm\,km\,s^{-1}\,Mpc^{-1}}$\fi}

\shorttitle{Profiles for [\ion{C}{2}] and Neon Emission Lines}
\shortauthors{Samsonyan et al.}

\begin{document}

\title{Neon and [\ion{C}{2}] 158 \um Emission Line Profiles in Dusty Starbursts and Active Galactic Nuclei} 

\author{Anahit Samsonyan\altaffilmark{1}, Daniel Weedman\altaffilmark{2}, Vianney Lebouteiller\altaffilmark{3}, Donald Barry\altaffilmark{2}, and Lusine Sargsyan\altaffilmark{2}} 

\altaffiltext{1}{Byurakan Astrophysical Observatory, Byurakan, Armenia, anahit.sam@gmail.com}
\altaffiltext {2}{Cornell Center for Astrophysics and Planetary Science, Cornell University, Ithaca, NY 14853; dweedman@astro.cornell.edu }
\altaffiltext {3} {Laboratoire AIM, CEA/DSM-CNRS-Universite Paris Diderot, DAPNIA/Service d'Astrophysique, Saclay, France}

\begin{abstract}
  
  The sample of 379 extragalactic sources is presented that have
  mid-infrared, high resolution spectroscopy with the \textit{Spitzer}
  Infrared Spectrograph (IRS) and also spectroscopy of the
  [\ion{C}{2}] 158 \um line with the \textit{Herschel} Photodetector
  Array Camera and Spectrometer (PACS).  The emission line profiles of
  [\ion{Ne}{2}] 12.81~\ums, [\ion{Ne}{3}] 15.55~\ums, and [\ion{C}{2}]
  158~\um are presented, and intrinsic line widths are determined
  (full width half maximum of Gaussian profiles after instrumental
  correction). All line profiles together with overlays comparing
  positions of PACS and IRS observations are made available in the
  Cornell Atlas of \textit{Spitzer} IRS Sources (CASSIS).  Sources are
  classified from AGN to starburst based on equivalent widths of the
  6.2~\um polycyclic aromatic hydrocarbon feature. It is found that
  intrinsic line widths do not change among classification for
  [\ion{C}{2}], with median widths of 207~km~s$^{-1}$ for AGN,
  248~km~s$^{-1}$ for composites, and 233~km~s$^{-1}$ for starbursts.
  The [\ion{Ne}{2}] line widths also do not change with
  classification, but [\ion{Ne}{3}] lines are progressively broader
  from starburst to AGN.  A small number of objects with unusually
  broad lines or unusual redshift differences in any feature are
  identified.

\end{abstract}


\section{Introduction}

Identifying and understanding the initial formation of massive
galaxies and quasars in the early universe is a fundamental goal of
observational cosmology.  A rapidly developing capability for tracing
luminosity sources to high redshifts is the observation of the
[\ion{C}{2}] 158~\um emission line at redshifts z $>$ 4 using ground
based submillimeter interferometers
\citep[e.g.][]{huy13,wan13,car13,ban15}, with detections now having
been made to z = 7.  This has long been known as the strongest
far-infrared line in most sources \citep{sta91, mal97, nik98, luh03,
  bra08}, often carrying $\sim$~1\% of the total source luminosity,
and is thought to be associated with star formation because it should
arise within the photodissociation region (PDR) surrounding starbursts
\citep{tie85,hel01,mal01,mei07}.

In many high redshift sources, especially dust obscured sources, the
[\ion{C}{2}] feature is the only diagnostic atomic emission line which
can be observed.  This makes it vital to optimize use of this line for
learning about intrinsic source properties.  Not only line
luminosities but also accurate line profiles are observed.  For
example, the submillimeter galaxy in the group BRI 1202-0725 at z =
4.7 has a [\ion{C}{2}] FWHM of 700~km~s$^{-1}$ measured with the
Atacama Large Millimeter Array, but the associated quasar has FWHM of
only 275~km~s$^{-1}$ \citep{car13}.  What can we learn from such
differences?

A great amount of diagnostic information concerning the observable
differences between dusty, obscured sources powered by active galactic
nuclei (AGN) and those powered by rapid star formation (``starbursts")
accumulated with the mid-infrared spectroscopy of the Infrared
Spectrograph (IRS; Houck et al.\ 2004) on the \textit{Spitzer} Space
Telescope \citep{wer04}.  Classifications of AGN and starburst sources
were developed using both the strength of PAH features in low
resolution spectra and various emission line ratios in high resolution
spectra
\citep[e.g.][]{bra06,far07,des07,ber09,wee09,vei09,tom10,wu10,sar11,sti13,ina13}.

More recently, numerous observations of the [\ion{C}{2}] line were
made \citep{sar12,far13,dia13,sar14,del14,dia14} using the
Photodetector Array Camera and Spectrometer (PACS) instrument
\citep{pog10} on the \textit{Herschel} Space Observatory \citep{pil10}
and included many sources that had previously been observed by the
IRS.  Our own previous studies were specifically designed to compare
the [\ion{C}{2}] line luminosities to mid-infrared polycyclic aromatic
hydrocarbon (PAH) emission features and mid-infrared emission lines
observed with the IRS in order to calibrate the star formation rate
(SFR) using the [\ion{C}{2}] line.  We found that [\ion{C}{2}]
luminosities most closely track [\ion{Ne}{2}] luminosities, from which
we concluded that the [\ion{C}{2}] is showing PDRs associated with the
HII regions seen in [\ion{Ne}{2}].

The observational study of forbidden emission line profiles has long
been a key technique for classifying and understanding active
galaxies.  The most extensive early studies used the optical
[\ion{O}{3}] line at 0.50~\ums, from which a conclusion was reached
that the width of this line could generally distinguish AGN and
starbursts \citep{fel82,whi85}.  All conclusions based on optical
lines were restricted to sources with little extinction by dust,
however, which led to uncertainties in interpreting differences among
profiles if different components of sources have different values of
extinction.  Now that forbidden emission line profiles are available
in mid-infrared and far infrared, comparisons can be made without this
uncertainty. An initial indication that such line profiles provide
interesting results is in the line profile comparisons by
\citet{spo13} among 24 ultraluminous infrared galaxies (ULIRGS) with
the goal of seeking outflows from AGN.

To evaluate further the utility of the [\ion{C}{2}] line for
understanding AGN and starbursts, we compare in the present paper the
[\ion{C}{2}] emission line profiles to the mid-infrared diagnostics
for all extragalactic sources which have archival observations
available for both PACS [\ion{C}{2}] and IRS spectra. This study of
line profiles is an extension of previous analyses of total
[\ion{C}{2}] line fluxes.  We present emission line profiles for
[\ion{C}{2}], [\ion{Ne}{2}] 12.81~\ums, and [\ion{Ne}{3}] 15.55~\ums
for extragalactic sources which have archival observations available
for both PACS [\ion{C}{2}] and IRS high resolution spectra.  Our
resulting sample of 379 extragalactic sources includes all sources
with both [\ion{C}{2}] and IRS high resolution observations, except
for nearby extended galaxies for which IRS observations were made in a
mapping mode.  The latter are omitted because our goal is to
understand overall, global properties of sources that would generally
be unresolved at high redshift. There are 390 [\ion{C}{2}] profiles in
our total sample, representing 379 different IRS pointings (the
remaining 11 are duplicate [\ion{C}{2}] observations made at the same
IRS position).

We present the source list and measurements of line widths and
redshifts along with overall comparisons among line widths.  All of
the [\ion{C}{2}] line profiles with Gaussian fits are made available
in the Cornell Atlas of \textit{Spitzer}/IRS Sources (CASSIS;
Lebouteiller et al.\ 2011, 2015\footnote{\url{http://cassis.sirtf.com}.  The
  [\ion{C}{2}] profiles are in
  \url{http://cassis.sirtf.com/herschel}. CASSIS is a product of the
  Infrared Science Center at Cornell University.}) Further comparisons
with other source properties will be considered in a future paper (in
preparation, Samsonyan 2016).

\section{Observations}

\subsection{Sample Selection}

Our sample was selected by examining abstracts of \textit{Herschel}
observing programs which described observations of extragalactic
emission line
sources\footnote{\url{http://www.cosmos.esa.int/web/herschel/observations}}
and then studying archival sources from these programs to identify
those for which the [\ion{C}{2}] line was observed, either in line or
range spectroscopy. Once all extragalactic sources with [\ion{C}{2}]
observations were identified, we searched for these sources in CASSIS
to locate sources having high resolution IRS observations, necessary
for line profile information.  The individual \textit{Herschel}
observing programs and the number of sources taken from each are
SDP-esturm-3, DDT-esturm-4, and KPGT-esturm-1(56); KPGT-smadde01-(13);
KPOT-pvanderw-1(11); GT1-lspinogl-4 and GT2-lspinogl-6(14);
OT1-dfarrah-1 and OT2-dfarrah-5(23); OT1-dweedman-1(112);
OT1-larmus-1(136); OT2-tdiazsan-1(4); OT1-lyoung-1(5);
OT1-nwerner-1(3); OT1-sveilleu-2(1); OT2-idelooze-1(1);
OT2-lsargsyan-1(2); OT2-nnesvadb-3(2); OT2-pguillar-7(5);
KPOT-aedge-1(1); and OT1-bweiner-1(1).  Although we intended a
complete search, it is possible that some sources having both
[\ion{C}{2}] and IRS high resolution observations have been
overlooked.  The source list and all results are given in Table~1.

\subsection{ IRS High Resolution Spectra}

A few previous studies have determined line widths from IRS high
resolution spectra \citep{das11,spo13, ina13}.  These results have all
been approximated by applying a uniform estimate of instrumental
resolution using the typical resolution of $\sim$ 500~km~s$^{-1}$ for
full width half maximum (FWHM) of the instrumental profile listed in
the instrument description.  For our analysis, we apply two additional
steps for improving determination of the instrumental profile.  The
first is to determine empirical instrumental profiles for individual
emission lines, and the second is to apply an ``optimal" extraction
for the high resolution spectra, which enhances signal to noise (S/N)
for unresolved sources.  These steps are described below.

Spectral extraction is the process of producing a one dimensional
spectrum from the two dimensional images obtained with the IRS
detectors.  The best possible extraction of unresolved sources applies
the point spread function (PSF) of the telescope/spectrograph
combination so that the pixels in the cross dispersion direction are
weighted by the fraction of source flux which falls on them, thereby
reducing background and instrumental noise.  This is defined as an
``optimal" extraction.  These extractions have been implemented for
the spectra shown in CASSIS using empirically determined PSFs for both
low resolution \citep{leb11} and high resolution \citep{leb15}.  If a
source is fully extended over the observing aperture, a better
alternative is a full aperture extraction that equally weights all
pixels within the aperture.  The choices provided in CASSIS for
spectral extractions of high resolution IRS spectra are described in
detail by \citet{leb15}.  All previous measures of IRS high resolution
spectra have used full aperture extractions, which are provided by the
``post-BCD" spectra of the \textit{Spitzer} Heritage
Archive\footnote{\url{http://irsa.ipac.caltech.edu/data/SPITZER/docs/spitzerdataarchives/}}.

For both low resolution and high resolution spectra, the source extent
is estimated in CASSIS by noting differences between the source
spatial profile compared to the PSF.  This estimate is more reliable
for low resolution spectra than for high resolution because of the
limited spatial profile sampling in high resolution; the PSF has full
width half maximum (FWHM) of about 2 pixels compared to the 6 pixel
length of the high resolution slits.  The majority of our sources are
not extended on the scale of the observing aperture, so our final
adopted measurements utilize the optimal extraction.  For our
objective of measuring line profiles, the best possible S/N is crucial
in order to obtain the best profiles.  This also leads to the choice
of optimal extraction even if some flux is underestimated for extended
sources.

The extraction labeled as ``optimal" in CASSIS uses a simultaneous
fitting to the two separate observing nod positions. This option also
provides an estimate of the underlying background by using residuals
beneath the fitted PSF.  Another alternative, ``optimal differential",
that differences the two nods before fitting the PSF profile is also
given in CASSIS for comparison to aid in removing bad pixels, but this
method gives a reliable flux calibration only for unresolved
sources. CASSIS illustrates both of the optimal extractions in
addition to a full aperture extraction so that users can make a final
choice but recommends ``optimal" if the source appears spatially
unresolved.

The IRS instrumental spectral resolution for the high resolution modes
is approximately defined by the two pixel projected width of the
observing aperture.  Spectral resolution varies among different lines
because the IRS high resolution spectrographs are echelle
spectrographs with different lines seen in different orders, so
resolution increases as the order increases.  Because our objective is
the best possible measurement of line widths, the instrumental
resolutions for the [\ion{Ne}{2}] and [\ion{Ne}{3}] lines presented in
this paper are measured empirically.  This is done using spatially
unresolved planetary nebulae in which these emission lines are strong.
The intrinsic widths of planetary nebula emission lines are much
smaller than the IRS instrumental line width so the observed line
profiles illustrate the instrumental profile.

The sources used are the planetary nebulae from the Large Magellanic
Cloud, Small Magellanic Cloud, and Galactic Halo included in
\citet{ber09b} and \citet{pot10}.  Because these planetaries are
spatially unresolved, the CASSIS optimal extraction is used to measure
the lines.  Gaussian fits match the observed profiles well so we adopt
instrumental profiles that are Gaussians described by the FWHM of the
planetary nebula profiles.

The FWHM and dispersion among individual planetaries is
FWHM([\ion{Ne}{2}]) = 331 $\pm$ 35~km~s$^{-1}$ and FWHM([\ion{Ne}{3}])
= 377 $\pm$ 35~km~s$^{-1}$ .  The somewhat better resolution for
[\ion{Ne}{2}] 12.81~\um is expected, because this line is in
short-high order 16 whereas [\ion{Ne}{3}] 15.55~\um is in order 13.
These instrumental line widths were also measured using full aperture
extractions; the full aperture extractions give consistent but
slightly larger median widths compared to the optimal extractions,
371~km~s$^{-1}$ for [\ion{Ne}{2}] and 465~km~s$^{-1}$ for
[\ion{Ne}{3}].  These differences can be explained by small tilts of
the observing slits compared to the dispersion direction of the
spectra.  The optimal extractions give more weight to the center of
the slit where any line smearing from tilt effects is negligible, but
the full extractions equally weight flux near the edges, where slit
tilts would produce a larger shift of the dispersed image.

Because the optimal extraction is used for sources in this paper, we
adopt as the instrumental resolution profile (FWHM$_{\rm res}$) the
values of 331~km~s$^{-1}$ for [\ion{Ne}{2}] and 377~km~s$^{-1}$ for
[\ion{Ne}{3}], with an uncertainty of about 10\% for each.  The
intrinsic FWHM$_{\rm intrinsic}$ of the observed spectral lines in
sources are then determined as FWHM$_{\rm intrinsic}$$^{2}$ =
FWHM$_{\rm observed}$$^{2}$ - FWHM$_{\rm res}$$^{2}$.

Redshift measurement for sources depend on the position of the line
centroids, which depends on the wavelength calibration of the spectra
(``WAVESAMP" files).  This calibration is applied differently in
optimal and full aperture extraction. To determine any systematic
uncertainties arising from the extraction technique, we compared line
centroids for the planetary nebulae and find that the 1 $\sigma$
dispersion between full aperture and optimal extractions for the Neon
lines is $\pm$ 25~km~s$^{-1}$.  These results mean that redshift
uncertainties arising from different analysis techniques for the high
resolution spectra are only $\pm$ 0.00008.

The observed results for line widths from the CASSIS optimal
extractions are given in Table~1. All IRS emission lines, including
those obtained with the alternative extractions, can be examined
within CASSIS by source name or coordinate using the browse tools
provided without requiring any downloads of the spectra. The CASSIS
results provide the actually observed coordinate of the source, which
sometimes differs from the nominal coordinate corresponding to the
source name in other data bases.  In ``cluster" mode for IRS
observations, several independent sources may be included within the
same Astronomical Observation Request (AOR), and these sources are
sometimes components of the same galaxy.  For these reasons, it is
necessary to use the accurate coordinates from CASSIS for comparisons
to locate the PACS spaxel corresponding to the same position, and
these coordinates are those in Table~1.

To determine uncertainties in FWHM for line profiles depending on
extraction techique, we compared both full aperture and optimal
extractions for the [\ion{Ne}{2}] and [\ion{Ne}{3}] lines in all
sources.  Line profiles were measured using Gaussian fits to the
[\ion{Ne}{2}] and [\ion{Ne}{3}] emission lines with a first order
underlying continuum, fitting with the Spectroscopic Modeling Analysis
and Reduction Tool (SMART; Higdon et al.\ 2004, Lebouteiller et
al. 2010). The dispersions among the differences between full aperture
and optimal extractions for the same profiles are a measure of the
observational uncertainties that arise from independent fitting of
undersampled profiles with different Gaussians.  For both lines, these
dispersions ($\pm$~50~km~s$^{-1}$ for [\ion{Ne}{2}] and
$\pm$~46~km~s$^{-1}$ for [\ion{Ne}{3}]) are $\sim$~10\% of the
observed median line width (470~km~s$^{-1}$ for [\ion{Ne}{2}] and
512~km~s$^{-1}$ for [\ion{Ne}{3}]).

Dispersions are a larger fraction of the measured width for narrower
lines so the fractional uncertainty is greater for narrow lines.  This
effect is an additional reason to prefer optimal extractions for the
line measurements. In some cases, the observed FWHM from the best
Gaussian fit is smaller than the instrumental resolution, which cannot
be physically correct.  In these cases, the adopted intrinsic FWHM is
arbitrarily listed as 100~km~s$^{-1}$ and displayed at that value in
Figures so that these anomalous cases can be recognized. There is an
offset in measured FWHM such that the full aperture measures are
systematically larger, by 20~km~s$^{-1}$ for [\ion{Ne}{2}] and
70~km~s$^{-1}$ for [\ion{Ne}{3}].  We attribute this difference to a
similar effect as noted above for instrumental resolution - that
weighting all pixels of the tilted slit evenly in the full aperture
extraction leads to a smearing of the line in direction of dispersion,
which artificially increases the line width.

\subsection{\textit{Herschel} PACS [\ion{C}{2}] Spectra}

The PACS instrument \citep{pog10} simultaneously obtains spectra at
25~positions in a source with square apertures called ``spaxels".  If
all sources were spatially unresolved and all pointings were perfect,
direct comparisons of lines observed with IRS and PACS would be
straightforward.  Although neither condition is perfectly met for all
of these observations, both the IRS slit and a PACS spaxel observe
similar fractions of a source.  The IRS short wavelength, high
resolution observing slit for the Neon lines is 4.7$\arcsec$ wide,
designed to match approximately the FWHM of the telescope diffraction
profile at these wavelengths. The PACS spaxels are 9.4$\arcsec$ wide,
also nearly equal to the FWHM of the \textit{Herschel} diffraction
profile at the [\ion{C}{2}] wavelength. Because of this similarity in
fraction of a source observed by the two spectrographs, we compare
line profiles seen in the IRS slit to the [\ion{C}{2}] profile seen in
the single PACS spaxel that is closest in location to the IRS slit, as
illustrated in Figure~1.

A factor which might affect profile widths in spatially resolved
sources is the size of the emitting region observed by PACS.  These
sizes vary by a large amount because distances range from 0.05~Mpc to
1003~Mpc.  In Table~1, we list the projected sizes of the PACS
9.4\arcsec~ spaxel based on the angular size distances to the sources.
Redshift independent distances are used if these are listed in the
NASA/IPAC Extragalactic Database (NED).  If not, redshift derived
distances are determined from \citet{wri06} using H$_0$ = 71~\kmsMpc,
$\Omega_{M}$=0.27 and $\Omega_{\Lambda}$=0.73.  The distribution in
projected spaxel size among sources in Table~1 is shown in Figure~2.
The distributions illustrate that the PACS results refer primarily to
global properties of sources on observed scales of many kpc.

\begin{figure}

\figurenum{1}
\includegraphics[scale= 0.36]{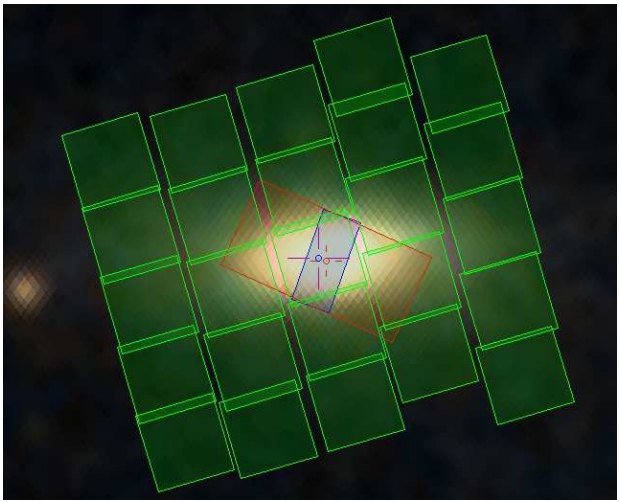}
\caption{Example of PACS spaxels compared to IRS high resolution
  slits, for source Markarian 18.  Squares are spaxels of size
  9.4\arcsec~ x 9.4\arcsec; large rectangle is IRS long-high slit, and
  small rectangle is IRS short-high slit. Crosses show coordinate of
  source in IRS slits. All Neon lines measured are in the short-high
  slit. The closest spaxel to position of the IRS slit is identified
  in Table~1; numbering convention used in Table~1 for rows and
  columns of spaxels is that upper left is 5,5; lower left is 5,1;
  upper right is 1,5; and lower right is 1,1. E is to the left and N
  to the top of the image. These overlays for all sources are shown in
  CASSIS at \protect\url{http://cassis.sirtf.com/herschel}.  }

\end{figure}

\begin{figure}

\figurenum{2}
\includegraphics[scale= 0.48]{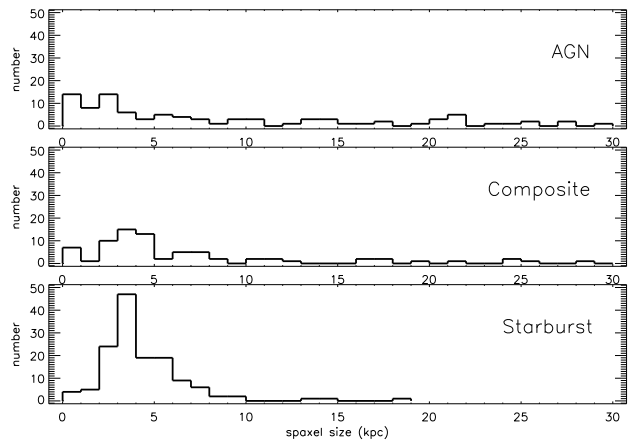}
\caption{ Distributions in projected size of 9.4\arcsec~ PACS spaxel
  among all sources in Table~1.  Histograms show actual number of
  sources of different classifications in each bin of width one kpc.}
\end{figure}

\begin{figure}

\figurenum{3}
\includegraphics[scale= 0.6]{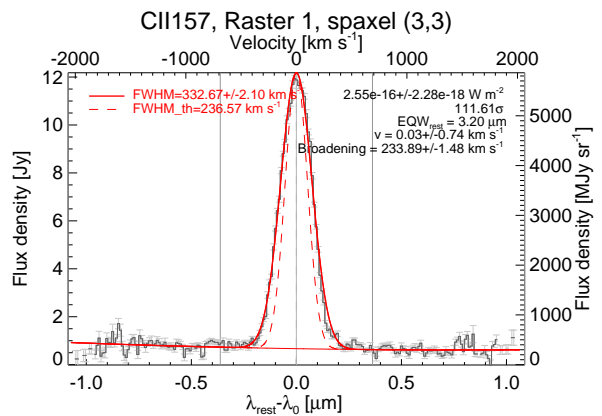}
\caption{Example of a [\ion{C}{2}] profile produced by PACSman and
  displayed in CASSIS, for source Markarian 18.  The fit by PACSman
  produces the FWHM and uncertainty of the observed Gaussian profile
  (solid line) fit to the data points (broken line and error bars),
  together with line flux and equivalent width. The line profile is
  shown at the [\ion{C}{2}] redshift given in Table~1, corrected to
  the local standard of rest.  Any assumed instrumental width
  (FWHM-th; dashed line) may be input to yield intrinsic width
  (Broadening) after removing this instrumental width from the
  observed profile. Instrumental FWHM used for all sources in this
  paper is 236~km~s$^{-1}$, determined empirically from observations
  of 30 Doradus, so intrinsic widths given in Table~1 are determined
  as FWHM$_{\rm intrinsic}$$^{2}$ = FWHM$_{\rm observed}$$^{2}$ -
  236$^{2}$.  All profiles used in this paper and the observed FWHM
  are illustrated in CASSIS at \protect\url{http://cassis.sirtf.com/herschel},
  although slightly different values of FWHM-th are sometimes shown
  for illustration.  For noisy profiles with an observed FWHM fit of
  $<$ 236~km~s$^{-1}$, an artificial value of FWHM-th is shown in the
  profiles so that the value determined for ``broadening" remains
  positive.  In Table~1, all profiles with observed FWHM $<$
  236~km~s$^{-1}$ are arbitarily assigned values of 100~km~s$^{-1}$
  for intrinsic width. }

\end{figure}

For our studies of line profiles, an important consideration is to
match the position of a PACS spaxel to the position of the IRS
aperture.  We did this individually for each source, producing
overlays of PACS spaxels and IRS aperture using the observed positions
given in the headers for the final IRS spectra, the coordinates in
Table~1.  Overlays such as the example in Figure~1 are shown in CASSIS
for all of our sources.  The individual spaxel corresponding most
closely to the position of the IRS aperture is listed in Table~1,
using the numbering convention shown in Figure~1, and the profile from
this spaxel is the [\ion{C}{2}] profile shown in CASSIS.

All [\ion{C}{2}] observations were reduced with version 12.1.0 of the
\textit{Herschel} Interactive Processing Environment (HIPE), together
with the ``PACSman" software \citep{leb12} to fit line profiles and
continuum within the spaxels. An illustration of a Gaussian line fit
provided by PACSman is in Figure~3, and all profiles which we use are
shown for the individual sources in CASSIS, searchable by source name
or coordinate. Displaying Gaussian fits is particularly useful for
seeking sources which might have a core Gaussian but broad or
asymmetric underlying wings \citep[e.g.][]{jan16}.  There are 34
sources in our sample for which [\ion{C}{2}] was observed but yields a
profile with insufficent S/N for an accurate width or radial velocity
measurement.  These profiles are shown as ``no data" in Table~1, but
they are illustrated in CASSIS with an attempted Gaussian profile fit
at the line position.

As was done for the IRS spectra, we determined an empirical
instrumental profile for the [\ion{C}{2}] observations.  This was done
using all 25 spaxels in the observation of the 30 Doradus HII region
listed in Table~1, because all spaxels give the same observed FWHM,
the intrinsic line widths are small compared to the instrumental
resolution, and the intrinsic line widths are known.  The observed
FWHM among these 25 profiles is 239~km~s$^{-1}$.  Internal motions
within the HII region have been measured for the optical [\ion{O}{3}]
0.50~\um line and have median FWHM of 45~km~s$^{-1}$ FWHM at spatial
resolutions similar to PACS \citep{sw72}.  Removing this intrinsic
width from the width observed by PACS results in an instrumental FWHM
of 236~km~s$^{-1}$.  The intrinsic FWHM$_{\rm intrinsic}$ of the
observed [\ion{C}{2}] lines in sources are then determined as
FWHM$_{\rm intrinsic}$$^{2}$ = FWHM$_{\rm observed}$$^{2}$ -
236$^{2}$.  The intrinsic widths are listed in Table~1.

\section{Analysis of [\ion{C}{2}] and Neon Line Profiles}

Our primary objective for the present paper is to present the
measurements in Table~1 and the profiles in CASSIS, which together
provide an observational atlas of [\ion{C}{2}] 158~\ums, [\ion{Ne}{2}]
12.81~\ums, and [\ion{Ne}{3}] 15.55~\um line widths that can be used
for comparing among various properties for a wide variety of sources.
Our eventual objective is to understand in more detail what determines
line widths in different galaxies.  For now, we only use these results
to determine if there are trends with classification as AGN or
starburst, or any systematic differences among [\ion{C}{2}] and
mid-infrared Neon lines.

\subsection{Comparisons of Line Profile Widths}

Previous studies using IRS line widths found some indications that
higher ionization lines sometimes show broader widths or outflow
effects from AGN, but these effects were found primarily for the high
ionization [SIV] and [OIV] lines in a few cases of sources with the
broadest lines \citep{das11,spo13, ina13}.  We emphasize the lower
ionization Neon lines, especially [\ion{Ne}{2}] because this line most
closely tracks [\ion{C}{2}] in comparisons of line fluxes
\citep{sar14}.  These Neon lines are also the best for comparison with
previous studies of line profiles of the strongest optical forbidden
line, [\ion{O}{3}] 0.50~\um \citep{fel82,whi85,nel96}, because of
similar ionization potentials (21.6~eV to produce \ion{Ne}{2}, 41.0~eV
to produce \ion{Ne}{3}, and 35.1~eV to produce \ion{O}{3}.)

The AGN/starburst classification used here, based on strength of the
6.2~\um PAH feature, is similar to that used in many previous studies,
although different authors use different PAH features and different
methods for measuring strength
\citep[e.g.][]{gen98,lau00,bra06,des07,vei09,tom10,wu10,sar11,pet11,sti13}.
The essential result used by all of these authors is that the PAH
features increase in strength as the starburst component increases
although different authors adopt somewhat different definitions of
AGN-dominated compared to starburst-dominated.  The quantitative
divisions illustrated in our results are that AGN have
EW(6.2~\ums)~$<$~0.1~\ums, composite AGN plus starburst have
0.1~\ums~$<$~EW(6.2~\ums)~$<$~0.4~\ums, and starbursts have
EW(6.2~\ums)~$>$~0.4~\ums.

The divisions we use began with an IRS spectroscopic survey of
starburst galaxies by \citet{bra06} and in a flux limited IRS survey
of sources including all classifications \citep{hou07}, subsequently
verified by comparison with optical classifications using several
hundred IRS spectra \citep{sar11}. Our primary motive for using the
6.2~\um feature instead of the stronger 11.3~\um feature has been to
allow classification of sources having sufficiently high redshifts
that the 11.3~\um feature is not visible in IRS spectra.  The
strongest PAH complex centered at 7.7~\um is not used because this
feature can be confused with an apparent spectral peak at similar
wavelength arising because of strong absorption on either side of the
peak in heavily obscured AGN. By illustrating various parameters
regarding line profiles as functions of EW(6.2~\ums), any trends with
AGN/starburst classes are readily seen.

The well established correlation between level of ionization and PAH
classification of AGN and starbursts is illustrated in \citet{sar14}
for various flux ratios among mid-infrared and [\ion{C}{2}] emission
lines as a function of PAH classification.  For our present use, the
important result is that the [\ion{Ne}{2}] line tracks [\ion{C}{2}]
independent of classification whereas the higher ionization
[\ion{Ne}{3}] line is stronger relative to [\ion{C}{2}] in AGN than in
starbursts. (Ratio of line fluxes for [\ion{Ne}{3}]/[\ion{Ne}{2}] for
the optimal extractions is given in Table~1 for each source.) This
implies that the line profiles from [\ion{Ne}{3}] should be more
affected by the AGN, whereas the [\ion{Ne}{2}] and [\ion{C}{2}] lines
should characterize the starburst component.  We first consider,
therefore, whether line profiles for these three features show any
differences between AGN and starbursts.

Comparisons of AGN/starburst classification with intrinsic line widths
(those given in Table~1, after correction for instrumental broadening)
are shown in Figures 4, 5, and 6.  We think that the most important
result is shown in Figure~4 for [\ion{C}{2}] because a primary goal of
our study is to seek any evidence that the [\ion{C}{2}] line width is
a diagnostic of AGN/starburst classification.  There is no evidence of
any trend, however. The medians and dispersions among [\ion{C}{2}]
line widths in Figure~4 are very similar for all classifications, with
medians 207~km~s$^{-1}$ for AGN, 248~km~s$^{-1}$ for composites, and
233~km~s$^{-1}$ for starbursts with dispersions in all cases of about
$\pm$ 130~km~s$^{-1}$. The most important result for [\ion{C}{2}] in
Figure~4, therefore, is that line widths do not change among
classification, indicating that [\ion{C}{2}] is dominated by the
starbursts within any source and not affected by the presence of an
AGN.  This confirms our conclusions in \citet{sar12} and \citet{sar14}
based only on line fluxes.

The results for [\ion{Ne}{2}] in Figure~5 show a similar result to
[\ion{C}{2}], with only a slight trend of increased widths for AGN;
the median FWHM for AGN and composites is 340~km~s$^{-1}$ compared to
300~km~s$^{-1}$ for starbursts.  The best comparison with [\ion{C}{2}]
is for the starbursts to rule out any broadening of [\ion{Ne}{2}] by
an AGN.  For starbursts, the difference between median FWHM of
233~km~s$^{-1}$ for [\ion{C}{2}] and 300~km~s$^{-1}$ for [\ion{Ne}{2}]
is interesting, but we cannot be confident that it is meaningful.
This difference is less than the uncertainty in the individual IRS
width measures shown in Figure~5.  The difference would disappear if
we assume a broader instrumental line width for [\ion{Ne}{2}] than we
have adopted, requiring an instrumental width of 380~km~s$^{-1}$
instead of the 331~km~s$^{-1}$ that is used.  This larger value is
close to the full aperture instrumental width discussed in section
2.2, but such a large instrumental width would not be appropriate for
unresolved sources, which require an optimal extraction. If this
larger instrumental width were used, full aperture extractions would
also have to be used, and these are systematically larger than the
optimal extraction widths that are given in Table~1.  The net result
would remain that intrinsic widths for [\ion{Ne}{2}] are greater than
those for [\ion{C}{2}]. We must conclude, therefore, that we cannot
precisely compare the [\ion{C}{2}] widths from PACS with the Neon
widths from the IRS because of uncertainties in determining the most
accurate extraction procedure for the IRS lines.

Despite these uncertainties among comparisons of [\ion{C}{2}] and Neon
line widths, the relative widths between the [\ion{Ne}{2}] and
[\ion{Ne}{3}] lines are meaningful. Unlike the [\ion{Ne}{2}] FWHM in
Figure~5, the [\ion{Ne}{3}] FWHM in Figure~6 show a trend for
increasing line widths from starbursts through AGN.  Both the median
values of line widths and upper values of the dispersions are
progressively larger from starbursts through composites to AGN, with
medians of 289~km~s$^{-1}$ for starbursts, 367~km~s$^{-1}$ for
composites, and 426~km~s$^{-1}$ for AGN. Upper values of the observed
one $\sigma$ dispersions are 417~km~s$^{-1}$ for starbursts,
580~km~s$^{-1}$ for composites, and 748~km~s$^{-1}$ for AGN.  This
trend for AGN to be offset to larger values of intrinsic [\ion{Ne}{3}]
line width is independent of the adopted value for [\ion{Ne}{3}]
instrumental width and demonstrates that some additional [\ion{Ne}{3}]
line broadening is associated with AGN. This is additional
confirmation that we should not expect similar profiles between
[\ion{C}{2}] and higher ionization features such as [\ion{Ne}{3}]
because they often arise in different physical regions of sources.

\begin{figure}

\figurenum{4}
\includegraphics[scale= 0.5]{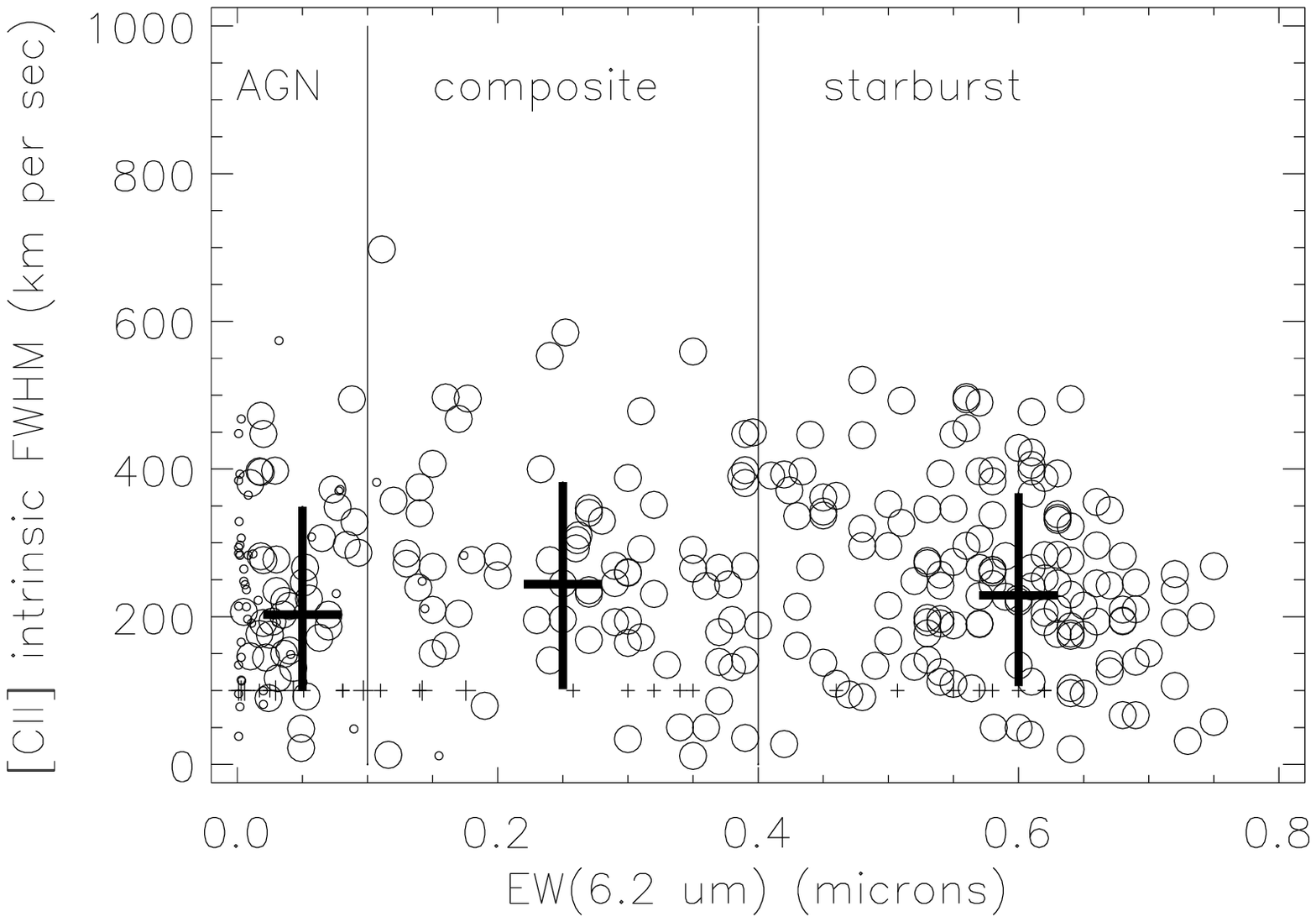}
\caption{Intrinsic FWHM of [\ion{C}{2}] 158~\um in~km~s$^{-1}$
  compared to EW(6.2~\ums) from Table~1. Large crosses show median
  line width and one $\sigma$ dispersion of line widths within
  classifications. Thin vertical lines divide classifications; AGN
  have EW(6.2~\ums) $<$ 0.1~\ums, composite AGN plus starburst have
  0.1~\ums $<$ EW(6.2~\ums) $<$ 0.4~\ums, and starbursts have
  EW(6.2~\ums) $>$ 0.4~\ums.  Small symbols are sources with measured
  line widths but upper limits in EW(6.2~\ums). Small crosses are
  sources in which the observed FWHM appears smaller than the
  instrumental resolution so FWHM arbitrarily assumed as
  100~km~s$^{-1}$. Observational uncertainty for [\ion{C}{2}] FWHM is
  typically smaller than size of symbols; uncertainties for each
  source are displayed in CASSIS.}
\end{figure}

\begin{figure}

\figurenum{5}
\includegraphics[scale= 0.5]{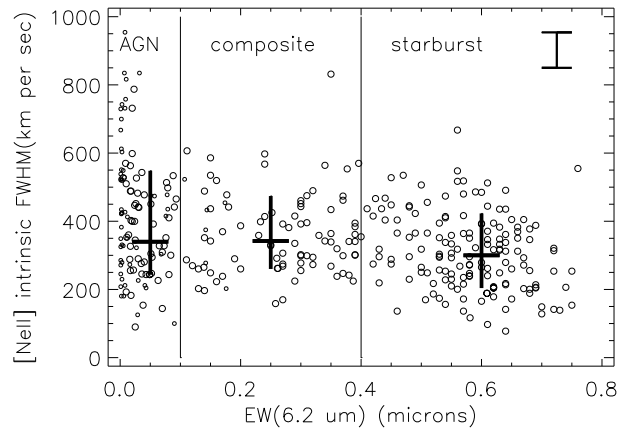}
\caption{Intrinsic FWHM of [\ion{Ne}{2}] 12.81~\um in~km~s$^{-1}$
  compared to EW(6.2~\ums) from Table~1. Symbols and explanations are
  as in Figure~4, with additional error bar that shows one $\sigma$
  dispersion of differences between full aperture and optimal
  extractions of the same line profiles which is the measure of
  uncertainty adopted for observed FWHM, discussed in section 2.2. }
\end{figure}

\begin{figure}

\figurenum{6}
\includegraphics[scale= 0.5]{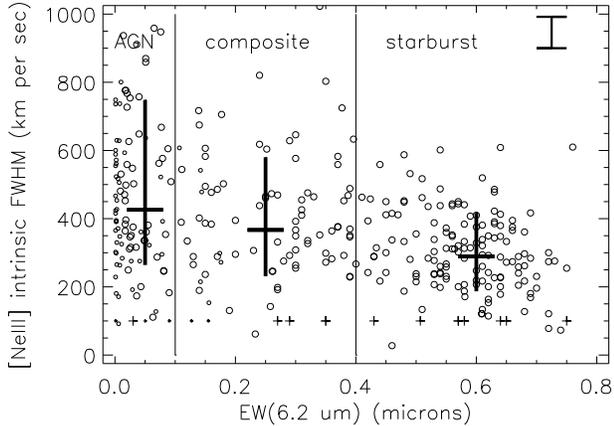}
\caption{Intrinsic FWHM of [\ion{Ne}{3}] 15.55~\um in~km~s$^{-1}$ compared to EW(6.2~\ums) from Table~1. Symbols and explanations are as in Figures 4 and 5.}
\end{figure}

Having confirmed that [\ion{C}{2}] and [\ion{Ne}{2}] line profiles
generally track one another, an additional use of the results in
Table~1 is to compare line widths for [\ion{C}{2}] and [\ion{Ne}{2}]
lines to find anomalous sources that are broad in one line but not the
other.  For example, are there sources where [\ion{C}{2}] arises in
regions where [\ion{Ne}{2}] is not seen, either because of too much
extinction or because there is a diffuse [\ion{C}{2}] not associated
with starbursts?  Or, are there sources in which even the low
ionization [\ion{Ne}{2}] is dominated by the AGN?  This comparison is
shown in Figure~7.  This Figure~is used to identify sources with
unusually broad features, defined as having FWHM that exceed the
median values by two $\sigma$. These sources are noted in Table~1 for
further future study.

A similar comparison is also made in Figure~8 between [\ion{C}{2}] and
[\ion{Ne}{3}], although it would be less surprising in this case to
find broad [\ion{Ne}{3}] associated with an AGN.  The few [\ion{C}{2}]
sources that equal or exceed the [\ion{Ne}{3}] width are particularly
interesting.  Broad profiles seen in this comparison are also noted in
Table~1.

\begin{figure}

\figurenum{7}
\includegraphics[scale= 0.5]{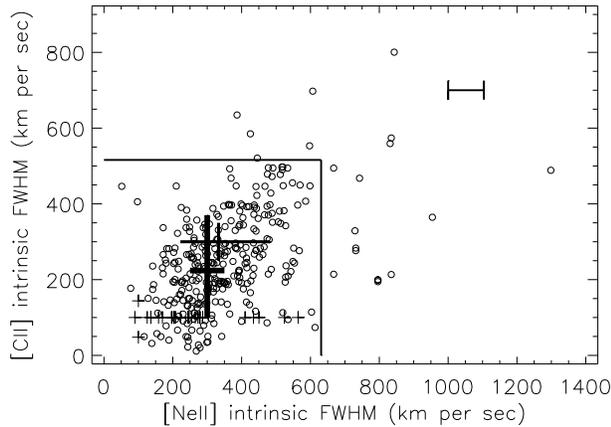}
\caption{Intrinsic [\ion{C}{2}] width compared to [\ion{Ne}{2}] width
  in~km~s$^{-1}$ from Table~1.  Small crosses are sources in which the
  observed FWHM appears smaller than the instrumental resolution so
  FWHM arbitrarily assumed as 100~km~s$^{-1}$.  Thick cross shows
  median (228~km~s$^{-1}$) and one $\sigma$ dispersion for
  [\ion{C}{2}], and thin cross shows median (333~km~s$^{-1}$) and one
  $\sigma$ dispersion for [\ion{Ne}{2}] using sources of all
  AGN/starburst classifications.  Sources outside of box have FWHM
  that exceed medians by two $\sigma$. Observational uncertainty for
  [\ion{C}{2}] FWHM is typically smaller than size of symbols;
  uncertainties for each source are displayed in CASSIS.  Error bar is
  observational uncertainty of [\ion{Ne}{2}] FWHM, as in Figure~5. }

\end{figure}

\begin{figure}

\figurenum{8}
\includegraphics[scale=0.5]{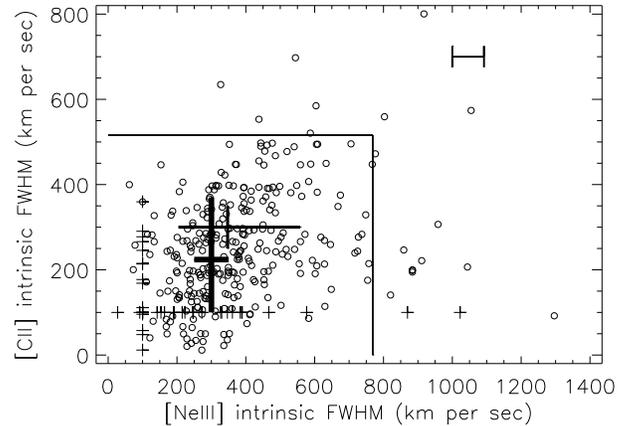}
\caption{Intrinsic [\ion{C}{2}] width compared to [\ion{Ne}{3}] width
  in~km~s$^{-1}$ from Table~1.  Symbols and explanations are as in
  Figure~7, with error bar showing observational uncertainty of
  [\ion{Ne}{3}] FWHM from Figure~6. }

\end{figure}

\subsection{Comparisons of Radial Velocities}

An alternative way to seek sources with significant differences
between the physical locations of [\ion{C}{2}] and Neon emission is to
compare observed radial velocities measured in these different
features.  These comparisons are in Figures 9 and 10.  The velocity
differences $dV$ between the profile centroids are given in Table~1,
determined from the redshifts measured independently for the
[\ion{C}{2}] and Neon lines and transformed to velocity units so that
$dV = c[z([\textrm{\ion{C}{2}}])-z(Neon)]$. Comparisons of $dV$ are made to the
[\ion{C}{2}] FWHM to learn if broader lines have systematically more
scatter in $dV$ measures, although we see no indication of this. From
these plots, sources with $dV$ differing from the medians by more than
2$\sigma$ are selected and noted in Table~1, because these sources may
also prove to be unusual.  Dispersions in $dV$ measures between
[\ion{C}{2}] and Neon are $\sim$ $\pm$ 50~km~s$^{-1}$, or $\sim$ 10\%
of the observed median FWHM for the Neon lines.

In both Figures 9 and 10, there is little difference in median radial
velocities determined from PACS [\ion{C}{2}] or IRS Neon.  For
$dV([\textrm{\ion{C}{2}}]-[\textrm{\ion{Ne}{2}}])$, the median difference is
21~km~s$^{-1}$ and is 34~km~s$^{-1}$ for
$dV([\textrm{\ion{C}{2}}]-[\textrm{\ion{Ne}{3}}])$.  These small systematic differences
are a confirmation of the careful wavelength calibrations that were
done independently for both instruments.  There is a known
instrumental effect in PACS which can lead to a skewed profile and
offset in velocity.  This effect arises when an unresolved source is
not perfectly centered in a spaxel
\footnote{\url{http://herschel.esac.esa.int/Docs/PACS/pdf/pacs-om.pdf},
  Figures 4.16 and 4.18} and can produce velocity offsets up to $\sim$
$\pm$ 30~km~s$^{-1}$ for [\ion{C}{2}].  This is less than the offsets
which we note as unusually large, and offsets caused by this
instrumental effect would be recognized by having a skewed profile.

\begin{figure}

\figurenum{9}
\includegraphics[scale=0.5]{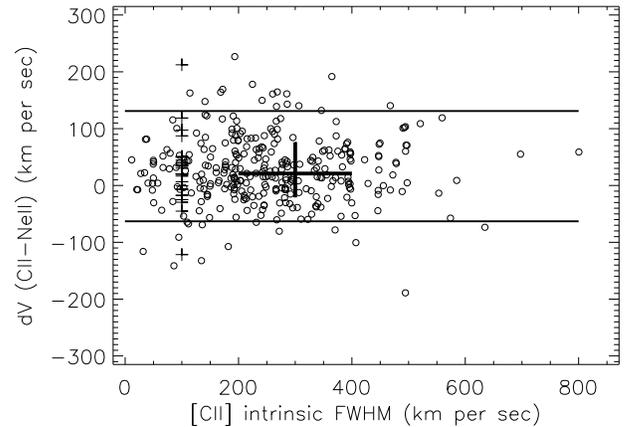}
\caption{Difference $dV$ between [\ion{C}{2}] radial velocity and
  [\ion{Ne}{2}] radial velocity in~km~s$^{-1}$ compared to
  [\ion{C}{2}] line width.  Large cross shows median (horizontal line)
  and 1 $\sigma$ dispersion about median (vertical line) of radial
  velocity differences.  Small crosses are sources in which the
  observed [\ion{C}{2}] FWHM appears smaller than the instrumental
  resolution so FWHM arbitrarily assumed as 100~km~s$^{-1}$. Sources
  outside of lines have $dV$ that exceed median by two $\sigma$.}

\end{figure}

\begin{figure}

\figurenum{10}
\includegraphics[scale=0.5]{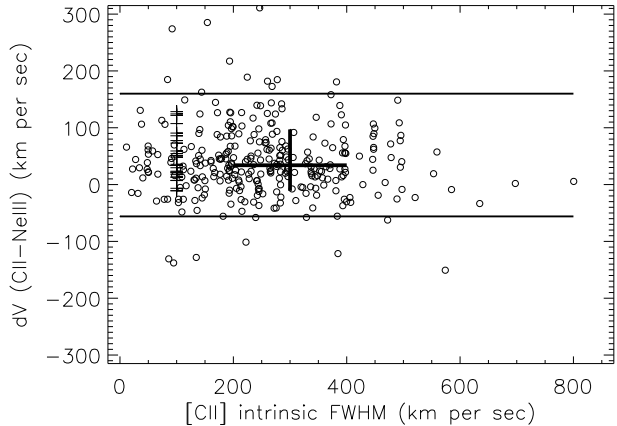}
\caption{Difference $dV$ in [\ion{C}{2}] radial velocity compared to
  [\ion{Ne}{3}] radial velocity in~km~s$^{-1}$.  Explanation of
  symbols same as in Figure~9. }

\end{figure}

\section{Summary and Conclusions}

We measure emission line profiles and redshifts for extragalactic
sources observed in both [\ion{C}{2}] 158~\um with \textit{Herschel}
PACS together with [\ion{Ne}{2}] 12.81~\um and [\ion{Ne}{3}] 15.55~\um
observed with the high resolution \textit{Spitzer} IRS. Data are
presented and compared for 379 different sources.  Results for
[\ion{C}{2}] and Neon are compared by locating the PACS observing
spaxel that most closely corresponds to the position of the IRS
slit. Intrinsic line profile widths are determined by applying
empirically measured instrumental widths from observed planetary
nebulae or HII regions.  All [\ion{C}{2}] and Neon line profiles
together with overlays of PACS spaxels compared to IRS slits are
illustrated in the CASSIS spectral atlas
(\url{http://cassis.sirtf.com/herschel}).

Sources are classified as AGN, composite, or starburst based on
equivalent width of the PAH 6.2~\um feature.  The median intrinsic
FWHM for [\ion{C}{2}] shows no change with classification, being
207~km~s$^{-1}$ for AGN, 248~km~s$^{-1}$ for composites, and
233~km~s$^{-1}$ for starbursts with dispersions in intrinsic line
widths of about $\pm$ 130~km~s$^{-1}$. Results show that [\ion{C}{2}]
line widths generally match those of [\ion{Ne}{2}], as previously
indicated in comparisons of line fluxes.  A small number of sources
are identified with unusually broad lines or with radial velocity
differences between [\ion{C}{2}] and Neon measures.  Accurate
redshifts are determined for sources as demonstrated by a systematic
difference of only 21~km~s$^{-1}$ between the independent measures of
[\ion{C}{2}] and [\ion{Ne}{2}] radial velocities using PACS and IRS.

\acknowledgments

This paper and the CASSIS atlas of IRS spectra are dedicated to our
friend and colleague James R. Houck, who died September 18, 2015.  Jim
Houck worked for 20 years to develop the Infrared Spectrograph for the
\textit{Spitzer} Space Telescope, which has enabled this research and
hundreds of other papers based on mid-infrared spectroscopy.  Thank
you, Jim.

We acknowledge use of archival data from the \textit{Herschel} PACS
instrument, which was developed by a consortium of institutes led by
MPE (Germany) and including UVIE (Austria); KU Leuven, CSL, IMEC
(Belgium); CEA, LAM (France); MPIA (Germany), INAF-IFSI/OAA/OAP/OAT,
LENS, SISSA (Italy); and IAC (Spain). This development was supported
by the funding agencies BMVIT (Austria), ESA-PRODEX (Belgium),
CEA/CNES (France), DLR (Germany), ASI/INAF (Italy), and CICYT/MCYT
(Spain). This research has made use of the NASA/IPAC Extragalactic
Database (NED) which is operated by the Jet Propulsion Laboratory,
California Institute of Technology, under contract with the National
Aeronautics and Space Administration. Partial support for this work at
Cornell University was provided by NASA through Contracts issued by
the NASA \textit{Herschel} Science Center. We thank Dieter Engels and
the Hamburger Sternwarte for hospitality during the initial phase of
this effort.

\newcommand\mytablenotetext[2]{\\$^{#1}${#2}}

\clearpage
\begin{landscape}
\LongTables
\tabletypesize{\scriptsize}

\begin{deluxetable}{lccccccccccccccc} 
\tablecolumns{16}
\tablecaption{Neon and [\ion{C}{2}] Line Widths and Redshifts}
\tablehead{
\colhead{No.}& \colhead{Herschel} &\colhead{Name\tablenotemark{a}} & \colhead{R.A.\tablenotemark{b}} & \colhead{Dec.\tablenotemark{b}} &\colhead{z\tablenotemark{c}} & \colhead{$dV[\textrm{\ion{Ne}{2}}]$\tablenotemark{d}}  & \colhead{$dV[\textrm{\ion{Ne}{3}}]$\tablenotemark{e}}& \colhead{FWHM\tablenotemark{f}} & \colhead{FWHM\tablenotemark{g}}  & \colhead{FWHM\tablenotemark{h}} & \colhead{EW\tablenotemark{i}} & \colhead{Neon\tablenotemark{j}} &\colhead{spax\tablenotemark{k}} &\colhead{size\tablenotemark{l}} &\colhead{ref\tablenotemark{n}} \\
\colhead{}& \colhead{AOR} & \colhead{} &\colhead{} &\colhead{} & \colhead{[\textrm{\ion{C}{2}}]} & \colhead{km~s$^{-1}$}  & \colhead{km~s$^{-1}$}  & \colhead{[\ion{C}{2}]}  & \colhead{[\ion{Ne}{2}]} & \colhead{[\ion{Ne}{3}]} & \colhead{\ums} & \colhead{ratio} &\colhead{} &\colhead{kpc} &\colhead{}
}
\startdata
1 & 1342235847 & Mkn334 & 00:03:09.62 & 21:57:36.60 & 0.022151 & 31 & 98 & 249 & 328 & 460 & 0.25 & 0.41 & 3.3 & 4.15  & 9 \\
2 & 1342235658 & NGC0023 & 00:09:53.36 & 25:55:27.70 & 0.015119 & 33 & -56 & 386 & 240 & 207 & 0.58 & 0.16 & 3.3 & 2.86  & 10 \\
3 & 1342199416 & NGC34/Mkn938 & 00:11:06.55 & -12:06:26.30 & 0.019418 & 52 & 73 & 364 & 448 & 450 & 0.45 & 0.14 & 3.3 & 3.65  & 6 \\
4 & 1342235846 & MCG-02-01-051/2 & 00:18:50.90 & -10:22:36.70 & 0.027243 & 13 & 34 & 197 & 321 & 300 & 0.72 & 0.22 & 3.3 & 5.08  & 9 \\
5 & 1342237570 & IRAS00188-0856 & 00:21:26.48 & -08:39:27.10 & 0.128447 & -12 & 92 & 310 & 328 & 959\tablenotemark{p} & 0.07 & 0.15 & 3.3 & 21.30  & 8 \\
6 & 1342237574 & IRAS00199-7426 & 00:22:07.01 & -74:09:41.70 & 0.096064 & 13 & 23 & 269 & 339 & 299 & 0.35 & 0.09 & 3.3 & 16.52  & 9 \\
7 & 1342199236 & Haro11 & 00:36:52.53 & -33:33:18.60 & 0.020717 & 50 & 38 & 157 & 249 & 236 & 0.15 & 3.65 & 3.3 & 3.88  & 5 \\
8 & 1342247543 & NGC185 & 00:38:57.97 & 48:20:14.60 & -0.000605 & \nodata & \nodata & 47 & \nodata & \nodata & 0.12 & \nodata & 3.3 & (0.03)  & 2 \\
\enddata
\end{deluxetable}\clearpage
\end{landscape}

\end{document}